\documentclass[a4paper]{aa}

\def\lsim{\lower.4ex\hbox{$\;\buildrel <\over{\scriptstyle\sim}\;$}} 
\usepackage{psfig,times} 
 \def\bib{\bibitem{}}

\def\start{\begin{itemize}}
\def\stop{\end{itemize}}
\def\apj{ApJ}

\begin{document}

\renewcommand{\vec}[1]{\mbox{\boldmath $#1$}} 
\def\bl{\par\vskip 12pt\noindent}  
\def\ea{et al.}                                         
\def\Pm{\mathop{\rm Pm}\nolimits}  
\def\qqq{\qquad\qquad\qquad}
\def\qq{\qquad\qquad}                      
\def\q{\qquad}  
\def\bib{\item}  
\def\beg{\begin{eqnarray}}  
\def\ende{\end{eqnarray}}  
\def\Om{{\it \Omega}} 
\title{Hydrodynamic stability in accretion disks under the combined  
       influence of shear and density stratification} 
\author{G\"unther R\"udiger\inst{1} \and Rainer Arlt\inst{1} 
\and Dima Shalybkov\inst{1,2}} 
\offprints{stability.tex} 
\institute{Astrophysikalisches Institut Potsdam, 
An der Sternwarte 16, D-14482 Potsdam, Germany \and
A.F. Ioffe Institute for Physics and Technology, 194021,
St. Petersburg, Russia} 
\date{\today} 
\abstract{The hydrodynamic stability of accretion disks is considered. 
The particular  question is whether the combined action of a (stable) vertical 
density stratification and a (stable) radial differential rotation gives rise
to a new instability for nonaxisymmetric modes of disturbances.
The existence of such an instability is {\em not} suggested by the 
well-known Solberg-H\o{}iland criterion. It is also not suggested by
a local analysis  for  disturbances in  general stratifications
of entropy and angular momentum which is presented in our Section 2
confirming the results of the Solberg-H\o{}iland criterion also for
nonaxisymmetric modes within the frame of ideal hydrodynamics but only in the frame of a short-wave
approximation for small $m$.  As a {\em necessary condition for stability} we 
find that  only {\em conservative}   external  forces are allowed to influence the stable disk.  As magnetic forces are never conservative, linear disk instabilities 
should only exist in the   magnetohydrodynamical regime which indeed contains the magnetorotational instability as a much-promising candidate.\\
To  overcome  some of the used approximations in a numerical approach,  the equations of the compressible adiabatic
hydrodynamics are integrated imposing initial nonaxisymmetric velocity perturbations
with  $m=1$  to   $m=200$. Only solutions with decaying kinetic energy
are found. The system always settles in a vertical  equilibrium
stratification according to pressure  balance with the gravitational
potential of the central object. 
}  
\authorrunning{G. R\"udiger,  R. Arlt  \& D. Shalybkov} 
\titlerunning{Hydrodynamic stability in accretion disks} 
 
\maketitle 
 
\section{Introduction: The Solberg-H\o{}iland criterion} 
Keplerian disks are, according to the Rayleigh criterion 
\beg 
{dj^2 \over dR} > 0 
\label{dj} 
\ende 
with $j=R^2 \Om$ being the angular momentum  
per unit mass and the distance $R$ from the rotation axis,  
hydrodynamically stable configurations. They are 
unstable configurations, however, under the influence of a (weak) magnetic
field if the disk plasma has a sufficiently 
high conductivity (Balbus \& Hawley 1991; Balbus 1995; Brandenburg et al. 1995;
Ziegler \& R\"udiger 2000). The question has now been formulated whether 
also other, nonmagnetic 
influences exist which in combination with the basic (negative) shear 
$d\Om/dR$ allow instabilities in the linear regime. We remember three 
recent discussions. Urpin \& Brandenburg (1998) consider the destabilizing 
action of any vertical dependence of $\Om$ on $z$ which may more than  
compensate the stabilization of the positive radial gradient of $j$. 
 
There is also the suggestion of Klahr \& Bodenheimer (2001) that
the {\em negative  radial} entropy 
stratification in thin Keplerian disks may act in the sense of a 
destabilization. Note that with the standard-alpha description of  
accretion disks one finds the positive value 
\beg 
{\partial S \over \partial R} \simeq {1\over 2} {C_v \over R}, 
\label{parts} 
\ende 
indicating, however, stability. 
Finally Richard et al. (2001) and Brandenburg \& Dintrans (2001) comment
on the idea that the 
vertical gradient of the density which always dominates accretion disk 
stratifications in combination with positive $\partial j/\partial R$ may 
destabilize the disk in the hydrodynamic regime. The extensive numerical
simulations in the local shearing box system have shown that Keplerian disks
(without vertical gradient of the angular velocity) are completely stable to
linear and nonlinear instabilities (Balbus et al. 1996). 
 
Most of the mentioned discussions are covered by the Solberg-H\o{}iland
conditions for dynamic {\em stability} which, for the rotation studied here,
take the form 
\beg 
{1\over R^3} {\partial j^2 \over \partial R} - {1\over C_p \rho} \nabla p \cdot 
\nabla S > 0, 
\label{stabi} 
\ende 
\beg 
{\partial p \over \partial z} \left({\partial j^2 \over \partial R} {\partial S 
\over \partial z} - {\partial j^2 \over \partial z} {\partial S \over \partial 
R}\right) < 0 
\label{partp} 
\ende 
(Tassoul 2000). If $\partial p/\partial z < 0$ (as is usual for accretion
disks), the latter relation turns into 
\beg 
{\partial j^2 \over \partial R} {\partial S \over \partial z} > {\partial j^2 
\over \partial z} {\partial S \over \partial R}, 
\label{partj} 
\ende 
which for Keplerian disks with  $\Om = \Om (R) \propto R^{-1.5}$ simply 
leads to the usual Schwarzschild criterion 
\beg 
{\partial S \over \partial z} > 0 
\label{partS} 
\ende 
for stability.  However, if the angular velocity $\Om$ depends on
the vertical  coordinate $z$ then Eq. (\ref{partj}) allows instability
in  the standard case of positive $\partial j/\partial R$  and 
$\partial S/ \partial z$ in case they are large enough.

Equation (\ref{stabi}) becomes 
\beg 
{1\over R^3} {\partial j^2 \over \partial R} + {g\over C_p} {\partial S \over 
\partial z} - 
{1\over \rho C_p} {\partial p \over \partial R} {\partial S \over \partial R} 
> 0 
\label{s3} 
\ende 
which with (\ref{partS}) and  for $dS/dR > 0$ ensures stability. 
It was adopted that $-g_z = g= \Om^2 z > 0$.
 
In the opposite case of a negative entropy gradient one at least needs 
\beg 
\left|{\partial S\over \partial R}\right| > {\rho C_p \Om^2 \over |\partial
p/\partial R|}   
\label{dS} 
\ende 
in order to violate Eq. (\ref{s3}) in the most optimistic case of vertical 
isentropy. 
 
If, on the other hand, the standard stratifications of accretion disks are $j = 
j(R)$ and $S=S(z)$, with $dj^2/dR > 0$  then the 
Solberg-H\o{}iland criterion reduces to 
\beg 
1+ {z\over C_p} {dS \over dz} > 0, \ \ \ \ \ \ \ \ \ {dS \over dz} > 0. 
\label{overcp} 
\ende 
(in the upper hemisphere). Obviously $dS/dz > 0$ is a {\em sufficient condition for stability of  Kepler disks 
after the Solberg-H\o{}iland criterion} also for the combined 
action of vertical density stratification and differential Kepler rotation.  
The same result have been derived in earlier papers by Livio \& Shaviv  
(1977), Abramowicz et al. (1984) and Elstner et al. (1989). The local  
analysis for ring-like disturbances by Elstner et al. 
demonstrates in all details that in the framework of an {\em ideal} 
hydrodynamics, i.e. for vanishing Shakura-Sunyaev alpha, the traditional 
Schwarzschild criterion $dS/dz > 0$ ensures stability for the Kepler  disk.  
If the disk can be considered as isothermal in the vertical direction then it follows  
\beg 
{dS \over dz} = - {\cal R\over \mu}  {d \log \rho \over dz} 
\label{overdz} 
\ende 
which is always positive for the typical density stratification, i.e. the disk should always be stable.  
  
\section{Combined stability conditions} 
In this Section  a local stability analysis of the equations of
ideal hydrodynamics in cylindrical  coordinates ($R,\phi,z$) is presented.
Sound waves and nonaxisymmetric disturbances are in particular allowed to
exist.

The three components of the momentum equation are 
\begin{eqnarray} 
\lefteqn{{\partial u_R \over \partial t} + u_R 
{\partial u_R \over \partial R} + {u_{\phi} \over R} 
{\partial u_R \over \partial \phi} +\,u_z {\partial u_R \over \partial z} -  
{ u_{\phi}^2 \over R}=}\nonumber\\  
&& \qqq = 
-{1 \over \rho}{\partial p \over \partial R} + g_R, 
\label{2}\\ 
\lefteqn{{\partial u_\phi \over \partial t} + u_R 
{\partial u_\phi \over \partial R} + {u_{\phi} \over R} 
{\partial u_\phi \over \partial \phi} + 
u_z {\partial u_\phi \over \partial z} + { u_{\phi} u_R \over R}=}\nonumber\\ 
&& \qqq = 
-{1 \over \rho R}{\partial p \over \partial \phi} + g_{\phi}, 
\label{3}\\ 
\lefteqn{{\partial u_z \over \partial t} + u_R 
{\partial u_z \over \partial R} + {u_{\phi} \over R} 
{\partial u_z \over \partial \phi} + u_z 
{\partial u_z \over \partial z} = 
-{1 \over \rho}{\partial p \over \partial z} + g_z.} 
\label{4} 
\ende 
The mass conservation reads 
\beg 
{\partial \rho \over \partial t} +  
{\partial \over \partial R}(\rho u_R) +  
{\rho u_R \over R} + 
{1 \over R} {\partial \over \partial \phi}(\rho u_\phi) +  
{\partial \over \partial z} (\rho u_z) = 0 
\label{1} 
\ende 
and the energy equation is 
\beg 
{\partial S \over \partial t} + u_R 
{\partial S \over \partial R} + {u_{\phi} \over R} 
{\partial S \over \partial \phi} + u_z 
{\partial S \over \partial z} = 0. 
\label{5} 
\ende 
As usual $\rho$ is the density, $p$ is the pressure, $S$ is the specific
entropy, $\vec{g}$ is the  vector of any acceleration. 
The  unperturbed state is assumed to be  
$u_{R,0}=u_{z,0}=0$, $u_{\phi,0}=R \Om(R,z)$, $\rho_0=\rho_0(R,z)$,
$p_0=p_0(R,z)$, $S_0=S_0(R,z)$, $g_R=g_R(R,z)$, $g_\phi=0$,
$g_z=g_z(R,z)$.
The only non-zero equations of the system (\ref{2})-(\ref{5})
are equations (\ref{2}) and (\ref{4}) which take the form
\beg 
g_R={1\over \rho_0} {\partial p_0\over \partial R}-R\Om^2, \q 
g_z={1\over \rho_0} {\partial p_0\over \partial z}.
\label{00}
\ende 
Finally, the ideal gas equation of state is used 
$p={\cal R}/\mu \rho T$, 
$dS=C_v d\log (p \rho^{-\gamma})$ with  
$C_v$ as the specific heat at constant volume, and  
$\gamma=C_p/C_v=5/3$.  
\subsection{Linearized equations}  
The perturbations in the force are neglected.
The perturbations $u_R, u_{\phi}, u_z, \rho_1, p_1$ to the basic state
are assumed small and the linearized system (\ref{2})-(\ref{5}) is 
\begin{eqnarray} 
%
%
\lefteqn{{\partial u_R \over \partial t} +  
\Om {\partial u_R \over \partial \phi}   
- 2 \Om u_{\phi}
+{1 \over \rho_0}{\partial p_1 \over \partial R}  
-{\rho_1 \over \rho_0^2}{\partial p_0 \over \partial R}=0,} 
\label{overt}\\ 
%
%
\lefteqn{{\partial u_\phi \over \partial t} + 
{1 \over R} {\partial \over \partial R} (R^2 \Om) u_R+ 
\Om {\partial u_\phi \over \partial \phi} + R {\partial \Om \over \partial 
z} u_z +}  \nonumber\\ 
&& \qq +{1 \over \rho_0 R}{\partial p_1 \over \partial \phi}=0, 
\label{overpart}\\ 
%
%
\lefteqn{{\partial u_z \over \partial t} +  
\Om {\partial u_z \over \partial \phi} +  
{1 \over \rho_0}{\partial p_1 \over \partial z} - 
{\rho_1 \over \rho_0^2}{\partial p_0 \over \partial z}=0,} \label{ovpartt}\\ 
%
%
\lefteqn{{1 \over \rho_0}{\partial \rho_1 \over \partial t} +  
{\partial u_R \over \partial R} +  
{u_R \over R} + {\partial u_z \over \partial z} + 
{1 \over R} {\partial u_\phi \over \partial \phi} + }\nonumber\\ 
&& \qq +{\Om \over \rho_0} {\partial \rho_1 \over \partial \phi} + 
{\partial \log\rho_0 \over \partial R} u_R + 
{\partial \log\rho_0 \over \partial z} u_z =0,\label{overrho}\\ 
%
%
\lefteqn{{\partial S_1 \over \partial t} + 
\Om {\partial S_1 \over \partial \phi} +  
u_R{\partial S_0 \over \partial R} + 
u_z{\partial S_0 \over \partial z}=0.} 
\label{partS1} 
\end{eqnarray} 
In the frame of our local approximation the equations are  
considered  only in small volume near some reference  
point ($R_0,\phi_0,z_0$). The coefficients 
in the equations are thus constant taken at  
($R_0,\phi_0,z_0$). As usual both the  
short-wave approximation and the small-$m$ approximation, i.e.   
\beg 
 |k_RR| \gg m, \ \ \ \ \ |k_zz| \gg m 
\label{krr} 
\ende 
are applied (cf. Meinel 1983).  
In this approximation  all perturbations can be expressed by  
the modal representation ${\rm exp}[{\rm i}(k_RR+m\phi+k_zz-\omega t)]$  
and the equations take the final form 
\begin{eqnarray} 
\lefteqn{-{\rm i}(\omega-m\Om)u_R - 2\Om u_\phi +{\rm i}k_R{p_1 \over \rho_0} 
-{\rho_1 \over \rho_0^2}{\partial p_0 \over \partial R}=0,} 
\label{ur}\\ 
\lefteqn{-{\rm i}(\omega-m\Om)u_\phi + {1 \over R}{\partial \over 
\partial R}(R^2 \Om) u_R+R {\partial \Om \over \partial z} u_z=0,} 
\label{uf}\\ 
\lefteqn{-{\rm i}(\omega-m\Om)u_z +{\rm i}k_z{p_1 \over \rho_0} 
-{\rho_1 \over \rho_0^2}{\partial p_0 \over \partial z}=0,} 
\label{uz}\\ 
\lefteqn{-{\rm i}(\omega-m\Om) {\rho_1 \over \rho_0}+ 
{\partial \log\rho_0 \over \partial R}u_R+  
{\partial \log\rho_0 \over \partial z}u_z\,+}\nonumber\\ 
&& \qqq \qqq +\,{\rm i}k_Ru_R+{\rm i}k_zu_z=0, 
\label{cont}\\ 
\lefteqn{-{\rm i}(\omega-m\Om) \left( {p_1 \over p_0} - 
\gamma {\rho_1 \over \rho_0} 
\right) +} \nonumber \\  
&& + {1 \over C_v} \left( {\partial S_0 \over \partial R} u_R+ 
{\partial S_0 \over \partial z} u_z \right) =0. 
\label{S} 
\end{eqnarray} 
Generally, it is 
\begin{eqnarray} 
{1 \over C_v }{\partial S_0 \over \partial R} = 
{\partial \over \partial R} \log \left( {p_0 \over \rho_0^\gamma} \right), 
\ \ \ \ \ \ 
{1 \over C_v}{\partial S_0 \over \partial z}= 
{\partial \over \partial z} \log \left( {p_0 \over \rho_0^\gamma} \right). 
\label{idS}
\end{eqnarray} 
The epicyclic frequency $\kappa$ and the  adiabatic sound speed
$c_{\rm ac}$ are 
\beg 
\kappa^2  = {1\over R^3} {\partial j^2\over \partial R}, \qq \q  
c_{\rm ac}^2 = \gamma {p_0 \over \rho_0}. 
\label{kappa} 
\ende 
The determinant of the above homogeneous fifth order  
system must vanish and the resulting dispersion relation  is  
\beg 
(\omega- m\Om) \cdot D =0
\label{mom}  
\ende
with
\beg
\lefteqn{D=(\omega-m\Om)^4-2 E (\omega-m\Om)^2 + F.}
\label{disp} 
\ende 
Here we have introduced the expressions
\beg
E={1\over2} \left[ (k_R^2+k_z^2)c_{\rm ac}^2 
+{1 \over \rho_0^2}{\partial p_0 \over \partial R} 
{\partial \rho_0 \over \partial R}+ {1 \over \rho_0^2} 
{\partial p_0 \over \partial z} 
{\partial \rho_0 \over \partial z} + \kappa^2 \right] \nonumber\\ 
\label{disp1} 
\ende 
and
\begin{eqnarray} 
\lefteqn{F= \left( {k_R \over \rho_0} {\partial p_0 \over \partial z} -  
{k_z \over \rho_0} {\partial p_0 \over \partial R} \right) \cdot }\nonumber\\ 
\lefteqn{\left( {k_R c_{\rm ac}^2 \over \gamma} 
{\partial \over \partial z} \log \left( {p_0 \over \rho_0^\gamma} \right) 
- 
{k_z c_{\rm ac}^2 \over \gamma} {\partial \over \partial R}
\log \left( {p_0 \over \rho_0^\gamma} \right) + \right. }\nonumber\\ 
\lefteqn{\left. +{{\rm i} \over \rho_0^2} \left({\partial p_0 \over \partial z} 
{\partial \rho_0 \over \partial R}-{\partial p_0 \over \partial R} 
{\partial \rho_0 \over \partial z} \right) \right) 
+ \kappa^2 \left(k_z^2c_{\rm ac}^2+ {1 \over \rho_0^2} 
{\partial p_0 \over \partial z}{\partial \rho_0 \over \partial z} 
\right)}\nonumber\\ 
\lefteqn{-R {\partial \Om^2 \over \partial z} \left(k_Rk_zc_{\rm ac}^2 
+{1 \over \rho_0^2} 
{\partial p_0 \over \partial z}{\partial \rho_0 \over \partial R}+
{\rm i} \left( {k_R \over \rho_0} {\partial p_0 \over \partial z} - 
{k_z \over \rho_0} {\partial p_0 \over \partial R} \right) \right).}\nonumber\\ 
\label{disp2} 
\end{eqnarray} 
For axisymmetric modes we simply have to 
replace  $\omega-m\Om$ by $\omega$. This procedure influences the  
frequencies rather than the stability. 

The stability of axisymmetric 
and nonaxisymmetric perturbations is thus defined by the same stability  
criterion in the approximation used.
The first factor in (\ref{mom}) describes the waves with
$\omega=m\Om$ and will not be considered below. The stability is
defined by the factor $D$. The roots of $D$ are 
\beg 
(\omega-m\Om)^2= E \pm \sqrt{E^2-F}. 
\label{roots}
\ende
According to (\ref{disp1}) the coefficient $E$ is real.
The flow is thus always unstable for negative $E$ and complex $F$.
Positive $E$ and real $F$ are, therefore, the {\it necessary conditions
for stability}.
According to (\ref{disp2}), $F$ is real if and only if
\beg
{1 \over \rho_0^2} \left( {\partial p_0 \over \partial z} 
{\partial \rho_0 \over \partial R} - {\partial p_0 \over \partial R}
{\partial \rho_0 \over \partial z} \right) - R {\partial \Om^2 \over
\partial z} = 0.
\label{cOm}
\ende
Inserting (\ref{cOm}) in (\ref{00}),
we find
\beg
{\partial g_R \over \partial z} = {\partial g_z \over \partial R}
\label{eg}
\ende
as the immediate consequence. Without forces this relation is always fulfilled.
Any {\em conservative force} is
a particular solution of (\ref{eg}).  If -- as it is in accretion disks -- the gravity 
balances the pressure and the centrifugal force, then Eq. (\ref{cOm}) is automatically fulfilled. 
Note that 
after the Poincare theorem for rotating media with potential force and
$\Om=\Om(R)$ both the density and the pressure can be written as
functions of the generalized potential so that (\ref{cOm}) is always fulfilled.

Generally, the magnetic field is {\em not} conservative and can never
fulfill the condition (\ref{eg}). This is the basic explanation for the
existence of the magnetorotational instability which is driven by (weak)
magnetic fields. 
\subsection{Sufficient conditions for stability}
Now we can suppose that the necessary condition (\ref{cOm}) is fulfilled. 

The flow is stable if both roots in (\ref{roots}) are real and
positive. The sufficient conditions for stability are therefore 
\beg 
E>0, \qq \q 0<F<E^2. 
\label{stab} 
\ende

Inserting (\ref{cOm}) into (\ref{disp2}) we find $F$ to be positive if
\beg
\lefteqn{ {k_R^2 \over k_z^2} N_z^2
+ {k_R \over k_z} {2 \over \gamma \rho_0} {\partial p_0 \over \partial z}
{\partial \over \partial R} \log \left( {p_0 \over \rho_0^\gamma} \right) 
+ N_R^2 + \kappa^2 +} \nonumber \\
&& \qq +{1 \over k_z^2 c_{\rm ac}^2 \rho_0^2}
{\partial p_0 \over \partial z} \left( \kappa^2
{\partial \rho_0 \over \partial z} - R {\partial \Om^2 \over \partial z}
{\partial \rho_0 \over \partial R} \right)>0,
\label{in1}
\ende
where
\beg
N_z^2=-{1 \over \gamma \rho_0} {\partial p_0 \over \partial z}
{\partial \over \partial z} \log \left( {p_0 \over \rho_0^\gamma} \right)
\label{Nz}
\ende
and
\beg
N_R^2=-{1 \over \gamma \rho_0} {\partial p_0 \over \partial R}
{\partial \over \partial R} \log \left( {p_0 \over \rho_0^\gamma} \right)
\label{nr}
\ende
are components of the Brunt-V\"ais\"al\"a frequency.
In our short-wave approximation (see (\ref{krr})) we can neglect the last
term in the left-hand side of (\ref{in1}). The expression (\ref{in1}) is thus a simple
quadratic expression in $k_R / k_z$. Two conditions will ensure its positivity: i) the expression is positive
for some value $k_R / k_z$ and ii) the expression has no real roots i.e.
the discriminant is negative.
The first condition is fulfilled if
\beg
N_R^2+N_z^2+\kappa^2>0.
\label{h1}
\ende
This is the Schwarzschild criterion for stability under
the presence of rotation. The second condition leads to
\beg
{\partial p_0 \over \partial z}
\left( \kappa^2
{\partial \over \partial z} \log \left( {p_0 \over \rho_0^\gamma} \right)-
R {\partial \Om^2 \over \partial z}
{\partial \over \partial R} \log \left( {p_0 \over \rho_0^\gamma} \right)
\right)<0.
\label{h2}
\ende
Equations (\ref{h1}) and (\ref{h2}) are exactly equivalent to the
Solberg-H\o{}iland conditions (\ref{stabi}) and
(\ref{partp}).

The Schwarzschild criterion (\ref{h1}) ensures
that $E>0$ and the short-wave approximation ensures that $F<E^2$ and
these relations do not yield any additional conditions. 

The Solberg-H\o{}iland conditions (\ref{stabi}) and (\ref{partp})
can also be obtained in the Boussinesq approximation. Goldreich \& Schubert
(1967) have already formulated the dispersion relation within the frame of
the Boussinesq approximation.
Nevertheless, our more general consideration allows to find
the new necessary condition for stability (\ref{cOm}).

\subsection{Limiting cases}

Let us consider two interesting limiting cases. 
Without rotation the Schwarzschild criterion (\ref{stabi}) takes the form
\beg
N_R^2+N_z^2>0,
\label{sch}
\ende
where $N_R$ and $N_z$ are given by (\ref{Nz}) and (\ref{nr}).
The stratification is thus always stable if  $p_0$ and $S_0$ have 
opposite stratifications -- which indeed is the standard case.
Nevertheless, Eq. (\ref{sch}) allows the stability also for the case
when $p_0$ and $S_0$ have the same stratifications in one direction
and opposite stratifications in another. Note, that according to
(\ref{sch}) the situation with pressure stratification without
any density stratification is always unstable.

Without density stratification the sufficient stability condition (\ref{stabi}) 
for a rotating flow is
\beg
\kappa^2> 
{1 \over \rho_0^2 c_{\rm ac}^2}
\left( {\partial p_0 \over \partial R} \right)^2 +
{1 \over \rho_0^2 c_{\rm ac}^2}
\left( {\partial p_0 \over \partial z} \right)^2.
\label{ray}
\ende
This is the classical Rayleigh criterion (\ref{dj}) for stability generalized to
the compressible case. Obviously, the right-hand side is positive
and the criterion (\ref{ray}) is more restrictive than the standard one.

\section{Numerical simulations} 
The main shortcoming in our presentation is the  
short-wave assumption (\ref{krr}) under which all the above considerations
are only valid. Without this assumption, Fourier modes are no longer
solutions of the linearized equations and we have to switch to much more
complicated mathematics involving integral equations
(see R\"udiger \& Kitchatinov 2000). We prefer to present numerical
simulations for the stability of (isothermal) density-stratified Keplerian
disks under the influence of finite disturbances. Along this way also
finite-amplitude disturbances can be considered within the frame of
a nonideal hydrodynamics.   
 
The above analytical study has shown that stratified disks are stable; 
under the assumption of a small scale in perturbations compared with  
the disk dimensions. We present a few numerical integrations of the 
hydrodynamic equations for the unrestricted case. The computations are  
based on the work of Arlt \& R\"udiger (2001) which required a stable, 
stratified Keplerian disk as a prerequisite. We adapt these simulations 
with adiabatic evolution and relatively strong nonaxisymmetric 
perturbations. 
 
 
 
The setup for our three-dimensional simulations applies a  
global, cylindrical computational  
domain with full azimuthal range, dimensionless radii from 4~to~6, 
and a vertical extent of 2~density scale heights on average, 
which is $-1$ to $+1$ in dimensionless units.  
The ZEUS-3D code (developed by Stone \& Norman 1992a,b; Stone et al. 1992) 
was used for the integration of the hydrodynamics. The modifications 
to the code are very close to those given in Arlt \& R\"udiger (2001);  
the induction equation is naturally not integrated here. 
\beg 
\frac{\partial\rho}{\partial t}+{\rm div}(\rho {\vec u})=0, 
\label{pro} 
\ende 
\beg 
\rho \frac{\partial{\vec u}}{\partial t}+ \rho({\vec u}\nabla) {\vec u}
= - \nabla p + \rho\, \vec{g}, 
\label{rhot} 
\ende 
\beg 
\frac{\partial e}{\partial t}+{\rm div}(e{\vec u})=-p\,{\rm div}{\vec u}, 
\label{et} 
\ende 
where ${\vec u}$, $\rho$, $e$, and $p$ have the usual meanings of 
velocity, density, energy density, and pressure. The source of gravitation 
is that of a point mass $M=10^5$ at $R=0$. The simulation does not include  
self-gravity in the disk. The polytrophic exponent is $\gamma=5/3$. 
The original ZEUS code provides artificial viscosities for 
improved shock evolution; we have put these terms to zero here. 
The advection interpolation is the second order van-Leer scheme. 
 
The conditions for the vertical boundaries are stress-free; 
no matter can exit the computational domain in vertical direction 
and the vertical derivatives of $u_R$ and $u_\phi$ vanish. 
The radial boundaries are also closed for the flow, and the radial 
derivative of $u_z$ vanishes. For the azimuthal flow, we have to 
adopt a modified boundary condition though. A Keplerian rotation 
profile is maintained into the boundary zones, based on the last 
zone of the integration domain. If the first azimuthal velocity 
of the computational domain is denoted by $u_\phi^{(1)}$, we use 
\begin{eqnarray} 
  u_\phi^{(0)}\phantom{^{-}} &=& u_\phi^{(1)} \sqrt{R^{(1)}/R^{(0)}}\\ 
  u_\phi^{(-1)} &=& u_\phi^{(1)} \sqrt{R^{(1)}/R^{(-1)}.} 
\end{eqnarray} 
The outer radial boundary is treated accordingly for $u_\phi^{(n+1)}$ 
and $u_\phi^{(n+2)}$. The azimuthal boundary conditions are periodic.  
 
 
A rough representation of a stratified disk with radius-independent 
density scale-height, was used for the initial conditions. We leave  
this configuration for free development under the influence of the 
gravitational potential. The initial conditions also involve 
a non-axisymmetric velocity perturbation of the form 
\begin{eqnarray} 
u_z &=& A \sin m\phi,\\ 
u_R &=& A \cos (m+1)\phi. 
\end{eqnarray} 
The amplitude $A$ is -- in terms of the Keplerian velocity in the 
middle of the annulus, $U_{\rm K}$ -- about $7\cdot 10^{-4}U_{\rm K}$. 
We have run two models with $m=1$ and different resolutions and 
an additional integration with $m=10$ in order to test short-wave 
perturbations. The models are summarized in Table~\ref{tab1}. 
The initial configuration is isothermal with an energy density of 
\begin{equation} 
e=\frac{1}{\gamma(\gamma-1)} \,\rho\, c_{\rm ac}^2, 
\end{equation} 
where the sound speed $c_{\rm ac}$ is 7\% of the Keplerian velocity 
$U_{\rm K}$, and the polytrope exponent is $\gamma=5/3$. 
 
\begin{table} 
\caption{Overview of simulation configurations. The quantity 
$m$ denotes the azimuthal mode of the initial perturbation. 
\label{tab1}} 
\begin{tabular}{lll} 
\hline 
Model & Resolution & $m$ \\ 
\hline 
I     & $31\times 61\times 351$ & 1 \\ 
II    & $71\times 71\times 351$ & 1 \\ 
III   & $31\times 61\times 701$ & 10 \\ 
IV    & $31\times 61\times 701$ & 200 \\ 
\hline 
\end{tabular} 
\end{table} 
 
Figure~1 shows the evolution of the directional kinetic energies 
for the $z$- and $r$-components. A gradual decay of fluid motion 
is observed. Velocity fluctuations can also be measured in terms 
of the correlation tensor element $Q_{R\phi}$ which is non-dimensionalized 
with the average pressure giving the Shakura-Sunyaev parameter 
$\alpha_{\rm SS}$ (cf. Arlt \& R\"udiger 2001). Figure~2 shows the evolution of $\alpha_{\rm SS}$ during  
30~rotation periods at the inner disk radius. A clear relaxation  
of the flow is seen. The initially strong velocity fluctuations 
settle the correct density stratification for the gravitational 
potential of the point mass (central star). The fluctuations in 
vertical direction reach $\pm14$\% of the Keplerian velocity $U_{\rm K}$. 
Those in the radial direction even reach $-27$\%. They are thus not  
small at all, and even a nonlinear instability might be noticed 
if existing. 
 
\begin{figure} 
\psfig{figure=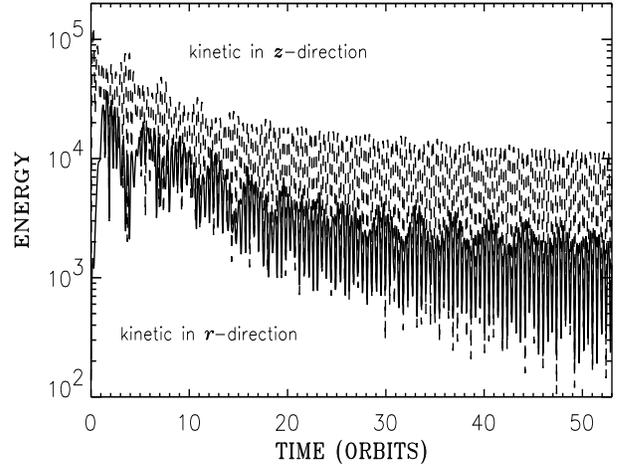,width=9cm,height=7cm} 
 \caption{Kinetic energies in the vertical and radial components of the 
flow from the run with $31\times 61\times 351$ grid points (Model~I).  
The solid line is the energy of the $r$-direction, the dashed 
line that of the $z$d-direction. Times are given in revolutions of  
the inner boundary of the computational domain} 
\label{energy} 
\end{figure}

\begin{figure} 
 \psfig{figure=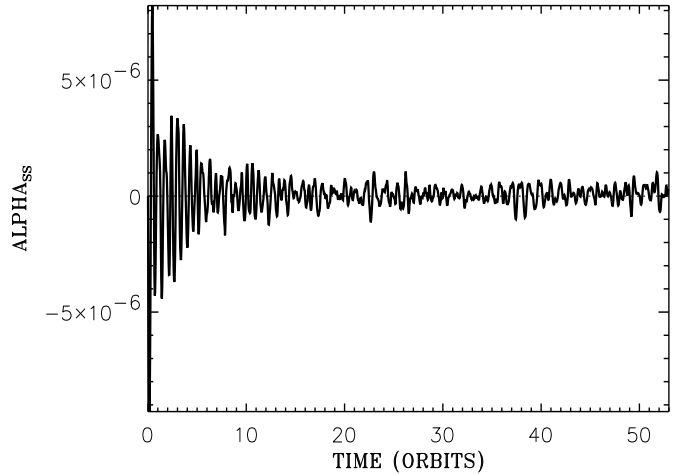,width=9cm,height=7cm} 
 \caption{Shakura-Sunyaev parameter $\alpha_{\rm SS}$ from the same 
model as in Fig.~1 (Model~I). Times are again 
in orbital periods of the inner boundary of the domain} 
\label{alphass} 
\end{figure} 
 
We can read an approximate decay time $\tau_{\rm D}$ from  
Fig.~1. On average, it amounts to about 20 orbital  
periods or $\tau_{\rm D}=3.18$ in non-dimensional 
units. As the main relaxation motions 
are vertical, we use the density scale-height for the length scale 
and obtain a viscosity of 0.03 caused by the finite resolution of 
the numerical grid. In terms if Reynolds numbers, we achieve 
\begin{equation} 
  {\rm Re} = \frac{U_{\rm in}\tau_{\rm D}}{H_\rho} = 1520 
\end{equation} 
near the inner edge of the annulus where the Keplerian velocity 
is $U_{\rm in}=158$. The density-scale height at the inner edge 
is $H_\rho=0.33$ in the initial configuration. 
 
A second run with significantly higher resolution -- in  
vertical direction in particular -- is shown in Fig.~3. 
No change in the decaying nature of the perturbation 
is found. 
 
An increase in azimuthal wave number of the perturbation does 
not provide instability either. Figure~\ref{m200} is the result 
of a simulation with a perturbation of very high wave number,
$u_r=A\sin 200\phi$, $u_z=A\cos 201\phi$, 
and doubled azimuthal resolution. The gas again settles towards an 
equilibrium state and very low kinetic energies in the radial and vertical
components. 
 
\begin{figure} 
\psfig{figure=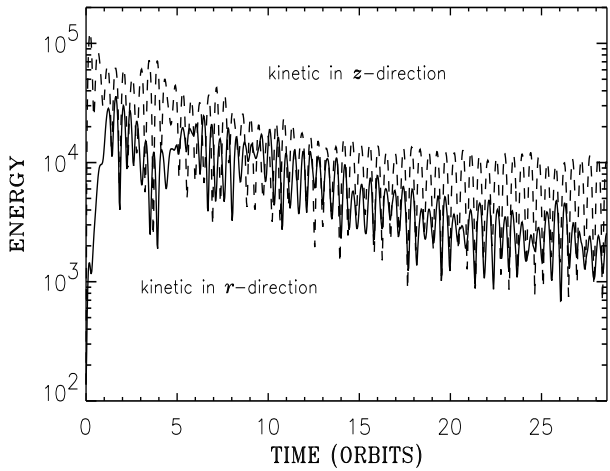,width=9cm,height=7cm}  
 \caption{Kinetic energies in the vertical (dashed) and radial  
(solid) components of the flow. The resolution of the model 
is $71\times 71\times 351$ grid points (Model~II). Times are again 
in orbital periods of the inner boundary of the domain} 
\label{high} 
\end{figure} 
 

\begin{figure} 
\psfig{figure=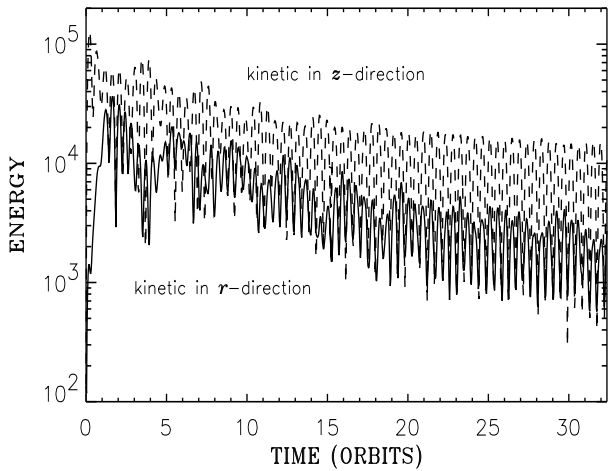,width=9cm,height=7cm}  
 \caption{Kinetic energies in the vertical (dashed) and radial  
(solid) components for the model with $m=200$ in the 
initial perturbation (Model~IV). The resolution of the computational  
domain is $31\times 61\times 701$} 
\label{m200} 
\end{figure} 
\section{Conclusions} 
We have shown in Section 1 that the Solberg-H\o{}iland conditions (\ref{stabi}) 
and (\ref{partp}) answer the question whether the combined action of stable (radial) 
shear and stable (vertical) density stratification can produce a new instability. 
The clear answer is No, but uncertainties remain about validity of the 
Solberg-H\o{}iland criterion for the case of nonaxisymmetric disturbances 
of the compressible accretion disk material.  Indeed, in Section  2 we were able to 
reproduce the Solberg-H\o{}iland criterion in the well-known formulation 
but only  with a short-wave approximation and  a small-$m$ approximation applied 
formulated in (\ref{krr}).  

Sound waves have been included into the consideration. Our dispersion relation (\ref{disp}) is of fourth order contrary to the dispersion relation of second order resulting in the Boussinesq approximation (Goldreich  \& Schubert 1967)  and contrary to the dispersion relation of fifth order resulting in nonideal hydrodynamics  (Abramowicz et al. 1984, 1990).   Along this way a new {\em necessary} condition for stability results which requires the external force which balances pressure and centrifugal force to be a conservative one, i.e. to possess a potential (cf. (\ref{cOm})). 

It is interesting to note that in the Boussinesq approximation for nonideal
fluids  the Solberg-H\o{}iland criterion changes to the
Goldreich-Schubert-Fricke criterion (Goldreich \& Schubert 1967; Smith \&
Fricke 1975; Urpin \& Brandenburg 1998) which no longer allows the existence  of a vertical shear for
stability, i.e. $\partial \Om/\partial z =0$.  And, indeed,  even in thin
accretion disks there is always  such a (weak) vertical shear $\partial
\Om/\partial z$ the meaning of which remains open. We have thus  numerically 
simulated the stability of such accretion disks using the ZEUS-3D code.  The
numerical  integration   of the nonlinear  adiabatic equations of the fully
compressible hydrodynamics with initially nonaxisymmetric velocity perturbations
does not lead to the onset of instability although  the disk, of course, does
possess a small vertical shear.  An overview about the simulated configurations
can be found in Table 1 which shows that we indeed have also considered  very
high values  of the azimuthal ``quantum'' number $m$ which do not fulfill the short-wave condition (\ref{krr}).

Based on our numerical simulations, we thus cannot support   findings 
 about a fast, hydrodynamic instability 
in stratified disk flows. 
Similar simulations including magnetic fields showed the  
magneto-rotational instability with growth rates near  
$0.5 P_{\rm orb}^{-1}$. Our hydrodynamic runs cover 30--50 orbits of 
the inner disk and show only decay in velocity fluctuations. 
Angular momentum transport hovers at $\alpha_{\rm SS}\lsim \pm10^{-6}$; 
the temporal average in Fig.~2 is $5\cdot 10^{-8}$.

\end{document}